\documentstyle[12pt]{article}

\begin{document}
\thispagestyle{empty}

\begin{center}
\LARGE \tt \bf{Non-Riemannian geometry of turbulent acoustic flows in analog gravity}
\end{center}

\vspace{1cm}

\begin{center} {\large L.C. Garcia de Andrade\footnote{Departamento de
F\'{\i}sica Te\'{o}rica - Instituto de F\'{\i}sica - Universidade do Estado do Rio de Janeiro-UERJ

Rua S\~{a}o Fco. Xavier 524, Rio de Janeiro, RJ

Maracan\~{a}, CEP:20550-003 , Brasil.

E-Mail.: garcia@dft.if.uerj.br}}
\end{center}

\vspace{1.0cm}

\begin{abstract}
Non-Riemannian geometry of acoustic non-relativistic turbulent flows is irrotationally perturbed generating a acoustic geometry model with acoustic metric and acoustic Cartan contortion. The contortion term is due to nonlinearities in the turbulent fluid. The acoustic curvature and acoustic contortion are given by Dirac delta distributions. Violation of Lorentz invariance due to turbulence is considered and analog gravity is suggested to be linked to planar acoustic domain walls. 
\end{abstract}      
\vspace{1.0cm}       
\begin{center}
\end{center}
\newpage
\pagestyle{myheadings}
\markright{\underline{Non-Riemannian turbulent flows}}
\section{Introduction}
\paragraph*{}
Investigation on analog gravity models in fluids and superfluids have been advanced in recent years by Visser and collaborators \cite{1} based on the basic reference of Unruh \cite{2} on the subject with the intention to investigate Hawking radiation of sonic black holes. Volovik \cite{3} also investigated analog models in the form of domain walls \cite{4} and turbulent flows. Recently Fischer and Visser \cite{1} have investigated the motion of phonons in the warped "spacetime" curved by analog gravity, in the form of Riemannian geometry of curved effective spacetimes of perfect nonrelativistic fluids. Also recently Schutzhold and Unruh \cite{5} investigated the newtonian gravity waves in water to investigate the sonic black holes in more detail. The comon feature of all these models is that dynamical equations are not Einstein gravitational equations but Euler and conservation of mass equations, which makes this sort of investigation dynamically easier than the perturbation problem in general relativistic cosmology. In this letter we investigate another sort of analog model based on non-Riemannian geometry of turbulent acoustic flows. This paper follows a previous papers on non-Riemannian geometry of rotational vortex acoustics \cite{6}. More recently this paper has been extended to allow for the presence of viscous fluids as well where then acoustic torsion is written in terms of vorticity of the viscosity of the fluid \cite{7}. In the present paper we go a step further and consider the gravitational analog of a acoustic turbulent flow and its corresponding acoustic metric and torsion. The effective geometry is written in terms of the turbulent flow speed of the background fluid while Cartan contortion is given as the product of the gradient of the turbulent speed by the irrotational perturbed velocity of the flow. This non-Riemannian effective geometry is shown to play the analogous role of domain walls in gravitational theory. The turbulent flow acts as a perturbation of flat effective spacetimes where turbulent flow speed appears in the Riemannian metric. The paper is organised as follows: In the section two we present the physics of the model as well as the acoustic metric of the turbulent fluid  layer. The acoustic Riemann tensor as well as the acpusric torsion are computed. Section 3 contains the computation of the dispersion relation for the turbulence and the acoustic Lorentz violation. Finally in the last section we discuss physical applications and future prospects. 
\section{The acoustic turbulent geometry}
The turbulent shear flow dynamical equations we consider here are given by the modified conservation equation 
\begin{equation}
\frac{D{\rho}}{Dt}+{\rho}{\nabla}.\vec{v}=q
\label{1}
\end{equation}
and the Euler equation
\begin{equation}
{\rho}\frac{D{\vec{v}}}{Dt}+{\nabla}p= \vec{f}
\label{2}
\end{equation}
here the source $\vec{f}$ is a dipolar source while the turbulence injection is given by $q=q(t,x,y)$. The operator $\frac{D}{Dt}$ reads 
\begin{equation}
\frac{D_{1}}{Dt}=(\frac{\partial}{{\partial}t}+U(x,y)\frac{\partial}{{\partial}x})
\label{3}
\end{equation}
Linearization of the dynamical equations of the flow is done according to the scheme
\begin{equation}
p= p_{0}+ p'
\label{4}
\end{equation}
\begin{equation}
{\rho}={\rho}_{0}+{\rho}'
\label{5}
\end{equation}
\begin{equation}
\vec{v}=\vec{U}+\vec{v'}
\label{6}
\end{equation}
where ${\nabla}{\rho}_{0}=0={\nabla}p_{0}$, whereas $c_{0}$ is the speed of sound in the medium and $\vec{U}=U(x,y)\vec{e_{1}}$ , produce the dynamical flow equations
\begin{equation}
\frac{D{\rho}'}{Dt}+{\rho}_{0}{\nabla}.\vec{v'}=q
\label{7}
\end{equation}
\begin{equation}
{\rho}_{0}(\frac{D_{1}\vec{v'}}{Dt})+({\nabla}U.\vec{v'})\vec{e_{1}}+{\nabla}p'= \vec{f}
\label{8}
\end{equation}
We shall assume throughout the paper that the flow is irrotational or $\vec{v'}={\nabla}{\psi}$ where ${\psi}$ is a scalar function. This assumption is reasonable since we shall next consider a shear thin layer and no bulk vorticity exists, which means that only vortices are local vorticity. We also assume that the external force can be obtained by the relation $\vec{f}= {\nabla}{\phi}$. The external force here is similar to what happens in the Newtonian gravitational analogue of gravity waves \cite{5}. After some usual algebraic manipulations one obtains
\begin{equation}
{\nabla}^{2}{\psi}-\frac{1}{{c^{2}}_{0}}\frac{{D_{1}}^{2}{\psi}}{D{t}^{2}}-({\nabla}U.\vec{v'})\vec{e_{1}}.\vec{v'}=\frac{1}{{\rho}_{0}}[q-\frac{D_{1}}{Dt}(\frac{{\phi}}{{c_{0}}^{2}})]
\label{9}
\end{equation}
By making use of expression (\ref{3}) into (\ref{1}) yields the generalised wave equation for the turbulent flow
\begin{equation}
{\nabla}^{2}{\psi}-\frac{1}{{c^{2}}_{0}}\frac{{\partial}^{2}{\psi}}{{\partial}{t}^{2}}-2U\frac{{\partial}^{2}{\psi}}{{\partial}x{\partial}t}-(\vec{e_{1}}.\vec{v'}){\nabla}U.{\nabla}{\psi}= \frac{1}{{\rho}_{0}}[q-\frac{D_{1}}{Dt}(\frac{{\phi}}{{c_{0}}^{2}})]
\label{10}
\end{equation}
where we have also drop out the term like $U^{2}$ , which is a reasonable assumption since we are dealing with a thin shear layer. Hence forth we shall supress the phonon speed $c_{0}$ by making it equal to one in analogy to what is done with the photon speed in vacuum in relativistic units. Throughout the calculations we also use the following relation
\begin{equation}
\frac{D_{1}{\rho}'}{Dt}= \frac{1}{{c^{2}}_{0}}\frac{D_{1}p'}{Dt}
\label{11}
\end{equation}
As was done in reference (\ref{6}) it is possible to extend the Unruh acoustic metric \cite{2} to an acoustic torsion by making use of the minimal coupling principle with torsion to write down the equation of a massless scalar field with source in terms of contortion vector $K^{i}=(K^{0},\vec{K})$, with $(i,j=0,1,2,3)$ as 
\begin{equation}
{\Box}{\psi}+ {K^{i}}{\partial}_{i}{\psi}= F
\label{12}
\end{equation}
where $F$ is the source term, the ${\Box}$ is the Riemannian D'Lambertian. Besides we have also considered a coordinate system where Cartan contortion is given by $K^{a}=(0,\vec{K})$. Thus by comparison with equation (\ref{11}) our $(2+1)$ dimensional effective spacetime case the generalised scalar wave equation (\ref{10}) can now be written as 
\begin{equation}
g^{ab}{\partial}_{a}{\partial}_{a}{\psi}+{\vec{K}}.{\nabla}{\psi}=[q-\frac{D_{1}}{Dt}{\phi}]
\label{13}
\end{equation} 
By considering a linear compact source as a Heaviside function $U=U_{0}yH(y)$ which is $U=U_{0}y$ when $-{\epsilon}< y<{\epsilon}$ ,and vanishes off this region, one can consider the approximation $U^{2}(y)=0$ which implies that the determinant of the metric is $g=det(g_{ab})= -(1+U^{2})= -1$. This in turn allows us to express the acoustic metric $g_{ab}$, where $(a,b=0,1,2)$ in $(2+1)$ effective spacetime dimensions in the shear flow region by the line element 
\begin{equation}
ds^{2}= -dt^{2}+dx^{2}+dy^{2}-2Udxdt
\label{14}
\end{equation}
By making use of the above approximation it is easy to compute the only nonvanishing component of the Riemann curvature is 
\begin{equation}
R_{0212}=\frac{1}{2}(\frac{d^{2}U}{dy^{2}})
\label{15}
\end{equation}
Since U can be expressed as a Heavised function the second derivative is a delta distribution and the laminar flow represents a topological planar defect. By comparing corresponding parts of equations (\ref{10}) and (\ref{13}) yields the acoustic contortion as
\begin{equation}
\vec{K}= (\vec{e_{1}}.\vec{v'}){\nabla}U
\label{16}
\end{equation}
Note that the acoustic contortion is given also by a Dirac delta distribution since the term ${\nabla}U(y)$ is a derivative of a Heaviside step function. It is interesting to note that in teleparallel theories Riemann curvature maybe expressed in terms of derivatives of torsion, therefore is not such a coincidence that acoustic curvature and contortion share the same sort of distributions. It is interesting to note that the Cartan contortion also depends on the gradient of turbulent speed. Metric (\ref{14}) is actually similar to the domain wall metric of the effective spacetime of Helium three \cite{3} given by Jacobson and Koike \cite{8}. Just to comparison with our metric we reproduce here their metric \cite{8}
\begin{equation}
ds^{2}=-(1-v^{2}c^{2}(x_{w}))dt^{2}+ 2vc(x_{w})^{-2}dtdx_{w}+c^{-2}(x_{w})d{x_{w}}^{2}
\label{17}
\end{equation}
where the main fundamental difference is that the domain wall function $c(x)$ depends on the texture inside the domain wall. Here the main fundamental difference is that the domain wall function $c(x)$ depends on the texture inside the domain wall.
\section{Acoustic Lorentz breaking in turbulent flows}
In this section we shall make use of the eikonal approximation to show that not only rotation and viscosity \cite{1} break acoustic Lorentz invariance , but also turbulence. Since would be more or less expected since many turbulent flows also possess rotation and viscosity, however this is not the case of our model here. Visser \cite{1} has shown that the viscous flow can break the acoustic Lorentz invariance by considering that the background fluid was at rest, however here our problem is much more involved since our background flow remains turbulent. To simplify matters let us now consider the wave equation in the situation where the mass injection q counterbalances the time derivative of the force potential ${\phi}$. In this case equation (\ref{10}) reduces to the sourseless wave equation
\begin{equation}
{\nabla}^{2}{\psi}-\frac{1}{{c^{2}}_{0}}\frac{{\partial}^{2}{\psi}}{{\partial}{t}^{2}}-2U\frac{{\partial}^{2}{\psi}}{{\partial}x{\partial}t}-(\vec{e_{1}}.\vec{v'}){\nabla}U.{\nabla}{\psi}= 0
\label{18}
\end{equation}
Taking the eikonal approximation
\begin{equation}
{\psi}= a(r)exp(-i[{\omega}t-\vec{k}.\vec{r}])
\label{19}
\end{equation}
into the turbulent wave equation (\ref{17}) one obtains the second order algebraic equation
\begin{equation}
-({\omega}-\vec{v}.\vec{k})^{2}+(c_{0}k_{x})^{2}+2Uk_{x}-(c_{0})^{2}(\vec{e_{1}}.{\vec{v'}}){\nabla}U.{\nabla}k_{x}=0
\label{20}
\end{equation}
Solving this equation for the frequency ${\omega}$ we obtain the following dispersion relation 
\begin{equation}
{\omega}_{\pm}= [\vec{v}.\vec{k}+k_{x}U]\pm\sqrt{(\vec{v}.\vec{k}+k_{x}U)^{2}- (\vec{v}.\vec{k})^{2}-{c_{0}}^{2}{k_{x}}^{2}+{c_{0}}^{2}(\vec{e_{1}}.\vec{v'}){\nabla}U.{\nabla}k_{x}}
\label{21}
\end{equation}
we note that here the turbulent speed contributes to the bulk frequency but also to the splitting of frequency and naturally to the Lorentz violation. Another intersting point that deserves attention is the fact that there is no dissipation due to turbulence.
\section{conclusions}
Non-Riemannian shear flow thin layer is proved to be similar to a moving domain wall when the turbulent flow is associated to an acoustic  curvature  constraint on the layer by Dirac delta curvature distributions. Cartan acoustic contortion is also given by a planar wall distribution. These acoustic turbulent flows may also be used to generated sound as recently showed by J. Barros \cite{9}.
\section*{Acknowledgement}
We are very much indebt to P.S.Letelier, H.P. de Oliveira, S. Bergliaffa, R. Schutzhold and J.D. Barros for helpful discussions on the subject of this paper, and to CNPq. and UERJ for financial support. 


\begin{thebibliography}{9}
\bibitem{1} U. Fischer and M. Visser,Warped spacetime for phonons moving in a perfect non-relativistic fluid,Europhysics Letters (2002). U. R. Fischer and M.Visser, Phys. Rev. Lett.88 , 110201(2002). M. Visser,Acoustic propagation in fluids:an unexpected example of Lorentz geometry, gr-qc?9311028.
\bibitem{2} W.G. Unruh, Phys. Rev. Lett. 46,1351 (1981).
\bibitem{3} G. Volovik, The Universe in a Helium droplet, Oxford University Press (2002). See also reprints from the Turbulence in Cosmology and Condensed matter workshop held in Finland in August 2004. 
\bibitem{4} L.C. Garcia de Andrade, Spacetime defects: Torsion walls, J. Math. Phys. (1998).
\bibitem{5} R. Schutzhold and W. Unruh, Phys. Rev. D 66,044019 (2002).
\bibitem{6} L.C. Garcia de Andrade, Phys. Rev. D 70,64004-1 (2004).
\bibitem{7} L.C. Garcia de Anddrade, Non-Riemannian viscous flows and acoustic Lorentz breaking, submitted to Phys. Rev. D (2004).
\bibitem{8} T. Jacobson and Keiko, Black hole and baby universe on a thin film of ${}^{3}He$,p.87 in M. Novello, M. Visser and G. Volovik,Artificial Black holes,Word Scientific (2002).
\bibitem{9} Jeanne de Barros,Dsc dissertation in portuguese.
\end{thebibliography}
\end{document}